\documentclass[aip,reprint]{revtex4-1}
\usepackage{graphicx}
\usepackage{dcolumn}
\usepackage{subfigure}
\usepackage{bm}
\usepackage{amsmath}
\usepackage{natbib}

\begin{document}

\preprint{}

\title{Pound-locking for characterization of superconducting microresonators} 



\author{T. Lindstr\"{o}m}


\author{J. Burnett}
\affiliation{National Physical Laboratory, Hampton Road, Teddington, TW11 0LW, UK}
\affiliation{Royal Holloway, University of London, Egham Hill, Egham, TW20 0EX, UK}
\author{M. Oxborrow}

\author{A Ya. Tzalenchuk}
\affiliation{National Physical Laboratory, Hampton Road, Teddington, TW11 0LW, UK}

\date{\today}

\begin{abstract}
We present a new application and implementation of the so-called Pound locking technique for the interrogation of superconducting microresonators.
We discuss how by comparing against stable frequency sources this technique can be used to characterize properties of resonators that can not be accessed using traditional methods. Specifically, by analyzing the noise spectra and the Allan deviation we obtain valuable information about the nature of the noise in superconducting planar resonators. This technique also greatly improves the read-out accuracy and measurement throughput compared to conventional methods.
\end{abstract}


\maketitle 

\section{Introduction}

Resonators are ubiquitous devices used in one form or another in most areas of physics and in many important applications, from filters in mobile phones to experiments in cavity Quantum Electrodynamics (QED)\cite{knight}. Over the past few years advances in on-chip quantum information processing have meant that planar microwave resonators have taken on new roles: as quantum buses\cite{simmonds20007}, for generating Fock states\cite{Hofheinz2008} and as read-out elements for solid-state qubits\cite{wallraff}. Superconducing resonators are also used as kinetic-inductance detectors\cite{Gao2007}. These applications involve measurements that need to be done quickly and accurately. It is natural to look at techniques already in use in the fields of optics and frequency metrology for inspiration on how to meet these new measurement challenges.  A commonly used technique for reading-out resonators in precision frequency metrology is Pound-locking\citep{Pound,Drever1983}, routinely used in quantum optics\footnote{Pound-locking is sometimes also referred to as Pound-Drever-Hall locking} and precision frequency metrology\cite{Rubiola} but surprisingly little known outside those fields.

The intrinsic properties of a resonator are described by its centre frequency $\nu_0$ and unloaded quality factor $Q_u=\nu_0/\Delta \nu$, where $\Delta \nu$ is the width of the resonance. This in turn is related to the attenuation rate  $\alpha=\pi \nu_0/2Q$. All these parameters are routinely measured using a vector network analyzer (VNA). Resonator measurements using a VNA are in principle accurate as long as the resonance shape is known and fitting techniques can be used to obtain Q and $\nu_0$, but are also quite slow, only measures the average behaviour and do not provide any direct information about the level of noise in the system. Another common approach is to use homodyne detection techniques where the carrier is mixed down to DC (using a mixer)\cite{Gao2007}. Pound-locking has a few important advantages over both these methods. The main improvement comes from the fact that the modulation frequency f$_m$ (see below) can be chosen to be much higher than the 1/f corner frequency of amplifiers etc; this is important since microwave amplifiers are very noisy at low offset frequencies, meaning the noise in a homodyne measurement\citep{Gao2007,Barends2009} scheme can be easily dominated by system noise. We typically use f$_m$ of 1-5~MHz, well above the corner frequency of our amplifiers. It is also a ''real time'' technique (within the bandwidth of the loop) in that it always "tracks" the centre frequency of the resonator irrespective of the magnitude of the fluctuation; leading to a significant speedup in the measurement throughput. Finally, it also allows for accurate measurements of very small shifts in the centre frequency of a resonator which is important in many experiments.

The purpose of this paper is to describe a new application and appropriate implementation of this well-established technique. We have adapted Pound-locking to the read-out and characterization of superconducting microresonators. We will also describe how one can obtain information about the nature of the noise in samples by using analysis techniques from precision frequency metrology. It is our hope that this work will expand the applications of the Pound locking technique.

\section{Pound-locking}
\begin{figure*}
\subfigure[The basic Pound loop]{
\includegraphics{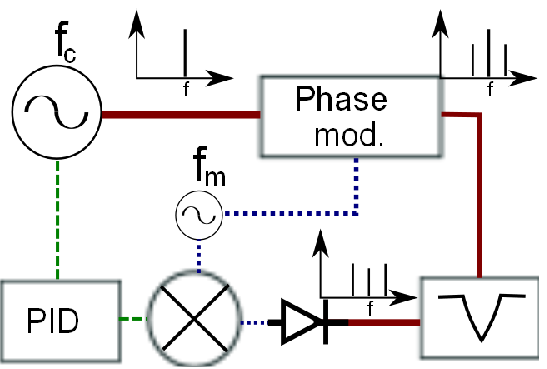}
\label{fig:pound}
}
\subfigure[Practical implementation of the Pound loop. Various auxiliary components (filters etc.) have been left out for clarity. ]{
\includegraphics[width=9cm]{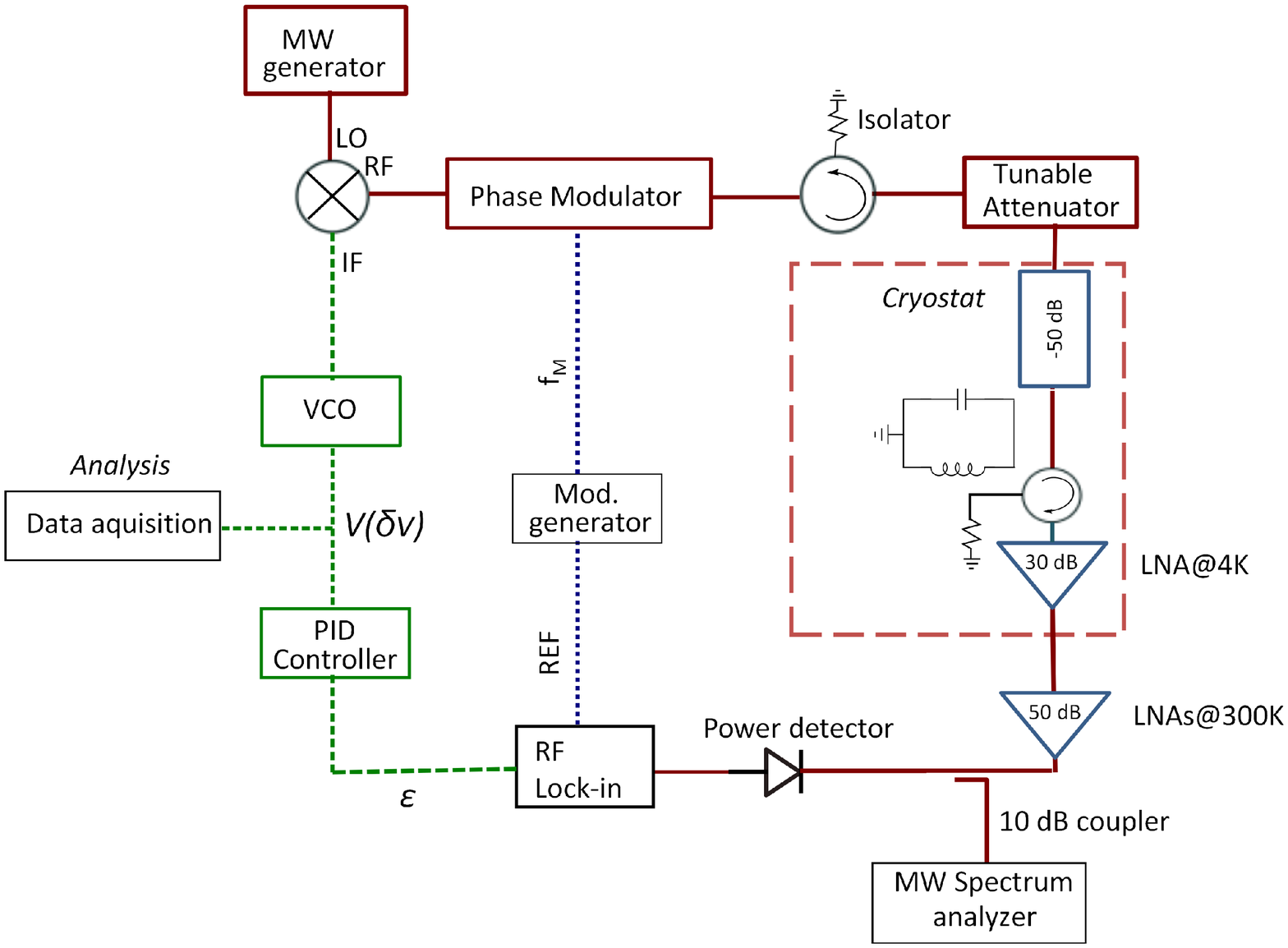}
\label{fig:setup}
}
\caption{\label{fig:setups}}
\end{figure*}

\begin{figure}
\includegraphics[width=8cm]{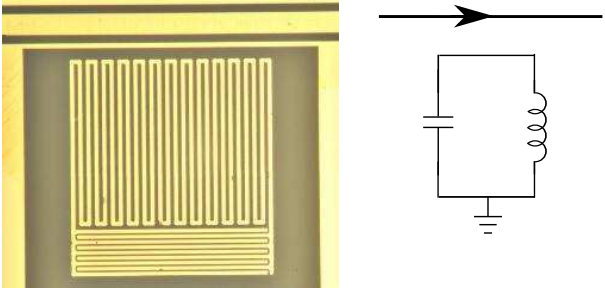}
\caption{Micrograph and equivalent circuit for one of our lumped element resonators. The electromagnetic radiation is
primarily inductively coupled to the inductive part of the resonator. The coupling strength is set by the mutual inductance between the coplanar transmission line and the resonator\label{fig:LE_res}.}%
\end{figure}

Figure \ref{fig:pound} show the basic schematic of a Pound-loop. Here we will describe how this is implemented in the case of microwave oscillators; for a very intuitive discussion of how Pound locking is used in optics see the review by Black\cite{black00}.
The main building blocks are a voltage-controlled oscillator (VCO) that generates a carrier of frequency $f_c$, a phase modulator used to add sidebands $\pm$f$_m$ to the signal, the resonator, a square-law power detector, a down-converting mixer and finally a regulator circuit which controls the VCO.
The principle of operation is as follows: assuming the carrier is near resonance and the modulation frequency has been chosen so that $f_m\gg \Delta \nu$ the sidebands do \textit{not} enter the resonator (note that this means that no extra power is injected into the resonator; nor is the bandwidth limited by $\alpha$).
The phase modulated signal going into the sample will be of the form \cite{black00}
\begin{equation}
V_{in}=V_0[J_0(\beta)+2jJ_1(\beta) \sin 2 \pi f_m t ] e^{2\pi j f_c  t}
\end{equation}
where $J_{0,1}$ are Bessel functions and $\beta$ is the modulation depth, it can be shown that the latter should be chosen so that the power in the sidebands is -3~dB smaller than the carrier. The resonator can be anything that has a "notch" type frequency response, most Pound loops are implemented using a circulator (meaning the sample is actually probed in reflection) but our planar resonator (see below and ref. \cite{ourprb09}) have this response in transmission, see fig. \ref{fig:LE_res}.
The scattering parameter $S_{21}$ for a piece of transmission line shunted by a resonator can be written
\begin{equation}
S_{21}=\frac{2(1+2j Q_u y)}{g+2(1+2j Q_u y)}
\label{eq:LEKIDfun}
\end{equation}
Where $g$ is a parameter determining the coupling strength and $y$ is the normalized centre frequency $(f-\nu_0)/\nu_0$.
After going through the sample the signal is  detected by a square-law diode after which it is de-modulated using a mixer.  The circuit therefore creates an error signal that is due to the interference between the transmitted signal and the energy leaking out from the resonator back into the transmission line. The signal from the diode can be written\citep{Anstie2006,Chang2000}
\begin{widetext}
\begin{multline}
\epsilon \propto 2V_0^2kJ_0(\beta)J_1(\beta)\{\text{Im}[S_{21}(f_c)](\text{Re}[S_{21}(f_c+f_m)]+\text{Re}[S_{21}(f_c-f_m)])-\\
\text{Re}[S_{21}(f_c)](\text{Im}[S_{21}(f_c+f_m)]+\text{Im}[S_{21}(f_c-f_m)])\}
\label{eq:errorsignal}
\end{multline}
\end{widetext}
where $\epsilon$ is the error signal; and $k$ is a constant that depends on the conversion efficiencies of the diode and the de-modulator.
\begin{figure}
\includegraphics[width=7.5cm]{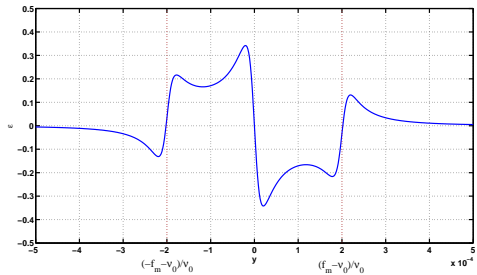}
\caption{Simulated typical error signal. The parameters are $f_m=\nu_0/2000$, Q=50000,g=2 and $\beta=1.08$.}%
\label{fig:errorsignal}
\end{figure}
The net effect of the Pound loop is that whenever the carrier is off resonance, the transmitted signal is phase shifted and will have an AM component, i.e. there is partial PM to AM conversion. The AM signal is then detected and used to correct the error. Hence, ideally this circuit will "lock" the frequency of the carrier $f_c$ to the resonance frequency $\nu_0$. Figure \ref{fig:errorsignal} shows what the error signal $\epsilon$ will look like for a loop with a typical resonator.
Once the system has locked to the resonator it will only undergo small excursions from $\nu_0$ and $\epsilon$ is nearly linear, using eq.\ref{eq:LEKIDfun} we can simplify eq.\ref{eq:errorsignal}
\begin{equation}
\epsilon_{lin} \propto kV_0^2 \frac{4gQ_u}{(g+2)^2}y
\end{equation}
where we have assumed optimum modulation depth ($\beta=1.08$) and coupling ($g=2$). Hence, the slope of the error signal depends on the uncoupled quality factor and the coupling strength of the resonator.

\section{Experimental}
The actual set up used in our experiment can be seen in \ref{fig:setup}. Our resonators are thin-film devices microfabricated on 5x10 mm$^2$ chips (see ref \cite{ourprb09} for a full description). The chip is glued to a copper cold-finger and bonded to a connectorized chip carrier which is enclosed in an aluminium box; the box is then mounted on the mixing chamber of a dilution fridge which has a base temperature of 30 mK. The in-signal to the resonator goes via a filtered and heavily attenuated (50~dB) line. The out-signal passes through an isolator and then into a low-noise cryogenic amplifier with a noise temperature of about 4~K.

The microwave signal (hereon referred to as the carrier) is generated by a voltage-controlled oscillator, in practice this is usually a microwave generator where the output frequency can be modulated using an external voltage (at a rate of for example 10 kHz/V). Another option is to use a local oscillator (LO) held at a fixed frequency and mix it with an intermediate frequency (IF) from a low-frequency VCO (in practice e.g. a function generator that can be frequency modulated) using for example a single sideband modulator~(SSB). The performance of the latter setup is in principle better since a low phase-noise LO can be used (the phase noise of the IF source is negligible) but makes the circuit more complicated (and care has to be taken to minimize the amplitude of the side-bands). Another advantage of this latter arrangement is that the frequency shift can be read out directly with high accuracy using a standard counter.
The carrier then passes through a phase modulator which adds sidebands $\pm f_m$. The shape of the error signal depends on the phase of the modulating signal, so this needs to be adjusted to compensate for lead/lag caused by the cables, filters etc.
The signal from the diode then goes to an RF lock-in amplifier (referenced to $f_m$) which is used with a very short time constant (as to not limit the bandwitdh of the loop), essentially being operated as a mixer with gain. The error signal in the loop is the in-phase signal from the lock-in amplifier, and is fed to the PID controller. The output voltage from the controller then controls the VCO (which in our case outputs the IF) in order to null the error; we generally use a deviation of 5-100 kHz/V.
The PID controller is configured so that it attempts to keep the error signal as close to zero as possible while still maintaining a loop bandwidth much larger than the frequency range of any noise we want to study. Care has to be taken to avoid dependence of the observed noise level on the loop parameters, this can  happen e.g. if the gain is too high (resulting in oscillations that distort the noise spectrum) or the bandwidth too low(resulting in a roll-off of the noise at higher frequencies) .

Most of our measurement data is acquired using a standard 16-bit data acquisition system connected to the output of the PID controller. Sampling rates from 10 Hz to about 50~kHz are used, depending on the measurement time which varies from a few seconds to tens of hours. We also monitor the output signal using  an FFT analyzer. In order to get an independent measure of the frequency we monitor the frequency of the IF source using a separate counter. All microwave equipment, counters etc in our experiment use an external 10~MHz frequency reference derived from a hydrogen maser.
Our current setup allows us to track any change to the resonance frequency of the resonator with a bandwidth of a few kHz (limited by the PID controller) and a "real-time" frequency resolution of a few parts in $10^8$ (limited by the intrinsic noise of the resonator).

Our experiment is essentially dual to that of the conventional setup used to improve the stability of an oscillator. In a conventional experiment a "bad" oscillator is locked to a "good" resonator, we are here doing the exact opposite. The frequency noise level of our resonator is much higher than that of the carrier; by observing how the circuit strives to null the error signal we obtain direct information about the resonator.  Moreover, since we are operating at offset frequencies smaller than $f_L=\nu_0/2Q$($\sim 75$~kHz), the Leeson effect\cite{Rubiola} will not affect the shape of the frequency spectrum \cite{rubiola_email}.

Our superconducting planar lumped element~(LE) resonators are  about 300x300 $\mu$m$^2$ in size, fabricated from niobium on a sapphire substrate. Several resonators (usually 5) are fabricated on the same chip and are coupled to the same transmission line allowing for frequency multiplexing.  The data shown here comes from a resonator with $\nu_0$ of 7.5~GHz and a loaded Q=50~000, it  is similar to the ones discussed in ref\cite{ourprb09}. Here we only show a subset of our results, for a full discussion see \cite{ourapl11}.

In order to verify that our data really represents properties of the sample and are not measurement artifacts; we also employ a number of reference samples. Here we will therefore also show data from a common dielectric resonator~(DRO) in the shape of a 'puck' about 7~mm in diameter mounted inside a metal box and measured at room temperature. In this experiment we locked to a mode with a resonance frequency of 5.12 GHz and a loaded Q=5500.
\section{Data analysis}

\begin{figure}
\includegraphics[width=8cm]{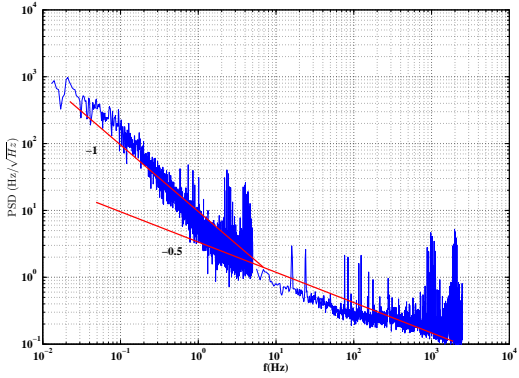}%
\caption{Root power spectral density (in units of $Hz/\sqrt{Hz}$) for a 7.5 GHz lumped element resonator measured at 30mK. the  solid lines have a slope of -1 and -0.5, respectivly and are meant as guides for the eye.\label{fig:PSD}}%
\end{figure}

The most straightforward way to measure noise is to use a spectrum analyzer or -equivalently- to calculate the power spectral density~(PSD) from sampled time-domain data. Figure~\ref{fig:PSD} shows a typical PSD from one of our LE resonators. Note that what is plotted is the frequency noise (in units of  $Hz/\sqrt{Hz}$)  meaning the PSD represent spectral information of how the centre frequency of the resonator changes.
The spectrum shows that random walk frequency noise (slope -1) dominates at low frequencies but that processes with a white (slope 0) spectra gradually takes over as the frequency increases.  The solid line with slope -0.5 can be identified with flicker frequency noise.

One limitation of PSD is that it is often very difficult to analyze ''slow'' processes. Another little known, but highly useful tool from the arsenal of frequency metrology is the Allan deviation (ADEV)\citep{NIST,Witt}. Here we have used the overlapping Allan deviation $\sigma$:
\begin{equation}
\label{eq:allandev}
\sigma^2=\frac{1}{2m^2(M-2m+1)} \sum_{j=1}^{M-2m+1} \left\{\sum_{i=j}^{j+m-1}(\overline{y}_{i+m}-\overline{y}_i) \right\} ^2
\end{equation}
where $M$ is the number of samples, $\tau_0$ is the measurement timebase and $\overline{y}$ is the time-averaged fractional frequency.
ADEV is the most common measure of time-domain stability, but other related types of measures such as the modified Allan deviation~(MDEV), Hadamard deviation etc are also very useful for analyzing and identifying noise (e.g. the MDEV in particular can distinguish between white and flicker phase noise).
Using the Allan deviation it is relatively easy to identify noise processes that happen over a long timescale. Just as in the case of the PSD we categorize the noise depending on the slope of the curve\cite{NIST} when plotted in a log-log plot. Table \ref{tab:table} shows a summary.
\begin{table}[b]
\caption{\label{tab:table} Noise processes and the resulting slope in the root PSD and Allen deviation(ADEV) plots}
\begin{ruledtabular}
\begin{tabular}{lcc}
Noise type & root PSD slope & ADEV slope\\
\hline
White phase & 1& -1\\
Flicker Phase& 0.5 & -1\\
White frequency& 0 & -0.5\\
Flicker frequency&-0.5& 0\\
Random walk &-1& 0.5 \\
Linear drift &-& 1 \\
\end{tabular}
\end{ruledtabular}
\end{table}
Note that we have included frequency drift in table \ref{tab:table}; although stricly speaking not a noise, drift due to e.g. temperature fluctuations are often seen in real measurements.

\begin{figure}
\includegraphics[width=8cm]{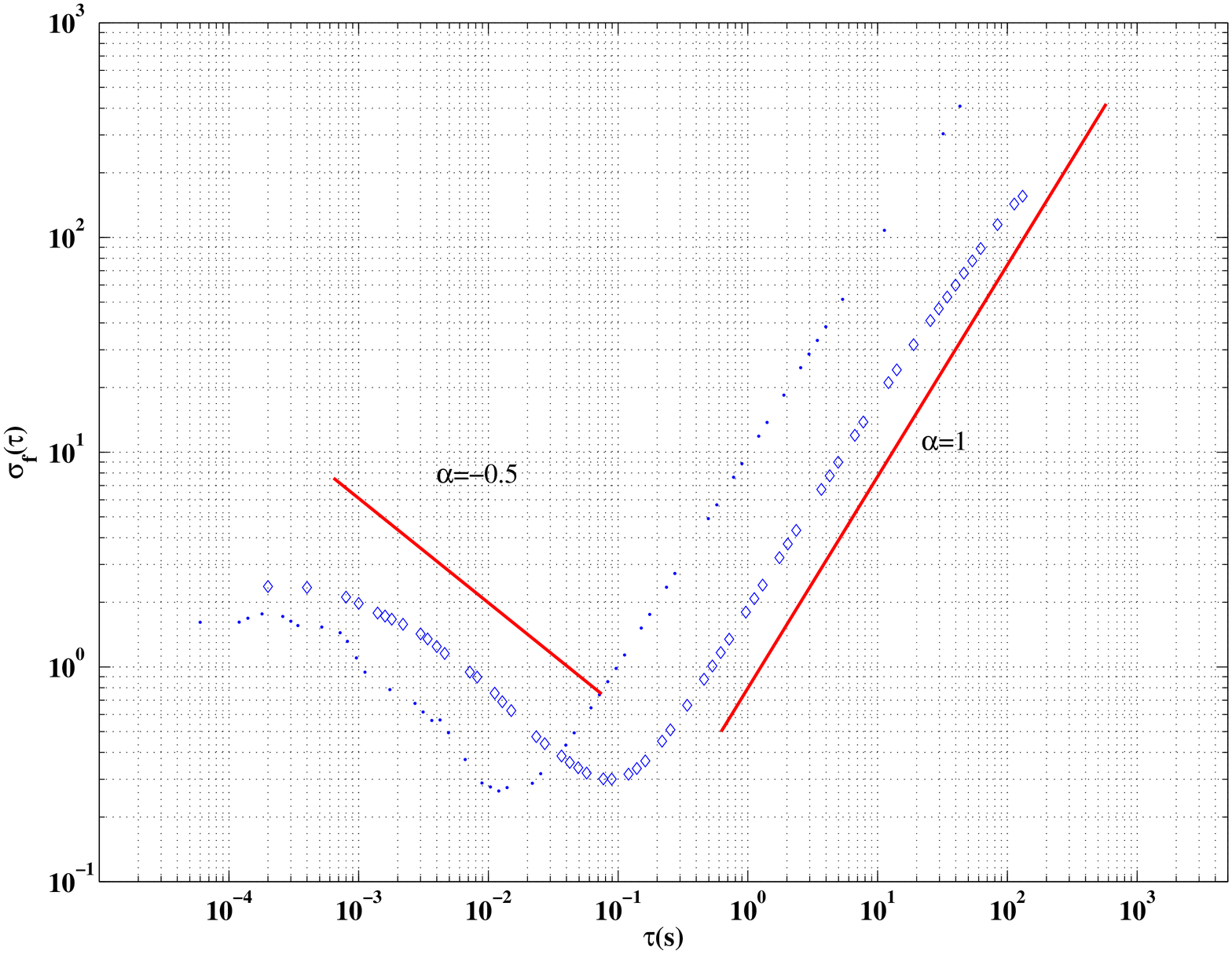}%
\caption{(colour online) Allan deviation plot for the dielectric resonator, with(diamonds) and without(points) temperature stabilisation\label{fig:diel_allan}.}
\label{fig:DRO}
\end{figure}

\begin{figure}
\includegraphics[width=8cm]{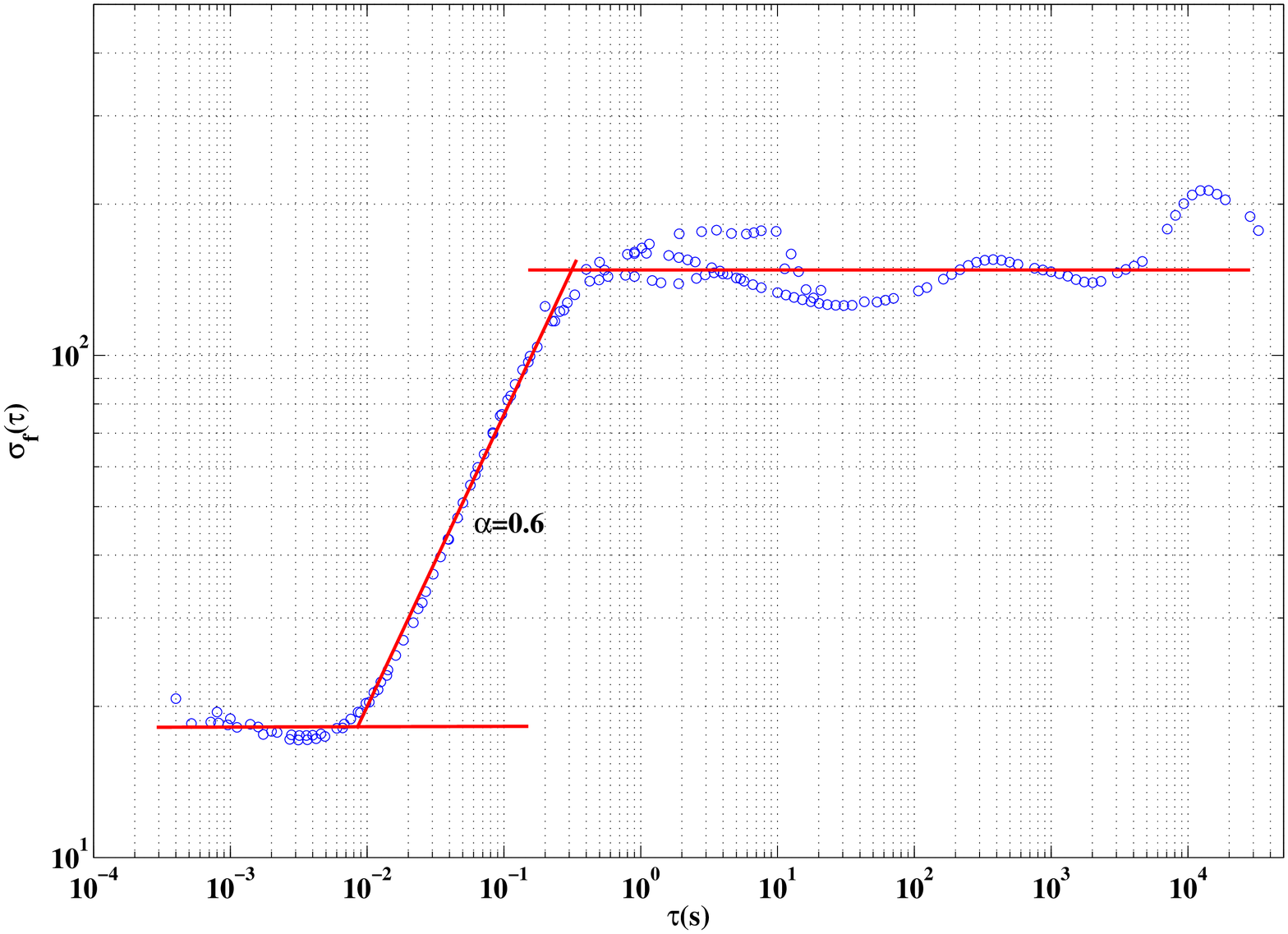}%
\caption{Allan deviation plot for a 7.5 GHz lumped element resonator measured at 30mK. The power in the resonator is approximately -100 dBm. \label{fig:res_allandev}}%
\end{figure}

In order to demonstrate how ADEV can be used to analyze the intrinsic noise of resonators we first show the experimental data from the DRO in figure \ref{fig:DRO}.
White frequency noise (slope -0.5)  dominates as expected for short times until linear drift (slope +1) takes over and the deviation increases rapidly.  This large drift in the resonance frequency (about 10-20 Hz/second)is due to a changing temperature and agrees with the specified temperature stability of a few ppm/K. By employing some rudimentary temperature stabilization we can shift the onset of the drift further to the right on the plot. According to this plot the best one can hope for is a short term stability (and read-out accuracy $\delta f$) of about 0.2 Hz; which agrees with the white noise level measured using an FFT analyzer as well as the direct read-out using a counter. This corresponds to a line splitting factor $\Delta \nu/\delta f$ of $5\cdot10^6$.
Figure \ref{fig:res_allandev} shows the ADEV -calculated using the same dataset as was used for the PSD- for the lumped element resonator measured at 30 mK. It is quite clear that the behaviour is very different from that of the dielectric resonator. Flicker frequency noise (slope 0) dominates at both long and short timescales, but the dominant feature is the sudden rise at timescales between 1 ms to 1s. We attribute this to random walk noise (slope 0.5).
It is interesting to compare this with the PSD shown in fig.~\ref{fig:PSD}. One obvious difference is that we can now distiguish the types of noise much more easily. The random walk noise -which is the most prominent feature in fig. \ref{fig:res_allandev}- is quite difficult to distinguish from other types of noise in the PSD and the ADEV also makes it easier to see over which timescale a particular noise process dominates.

Note that the line splitting factor at long times for the LE resonator is about 1500; much worse than for the dieletric resonator. This is simply a consequence of the fact that the lumped element resonator is -despite having a much higher Q- much worse at stabilizing the Pound loop. This in turn can be attributed to the fact that the LE resonator is subject to noise from its environment: the flicker frequency noise is most likely due to the presence of two-level fluctuators\cite{Gao2007}; the random walk noise might be due to random changes in the dielectric environment (due to hopping) or the movement of trapped vortices in the superconducting film.
\section{Conclusion}
By bringing well-known measurement techniques and analysis methods from time and frequency metrology to bear on the problem of the nature of noise in superconducting resonators one can gain a lot of information not readily obtainable using more widely known techniques. By implementing a Pound loop and analyzing the data using the Allan deviation we are able to identify noise processes that are hard to distinguish by conventional means. Moreover, the fact that the loop directly measures the resonance frequency makes the analysis more straigtforward.
Finally, we would also like to mention that the ability to "track" the value of $\nu_0$ can also lead to significant improvement in measurement throughput. A good example is measurements of $\nu_0$ vs. temperature which can e.g. be used to determine the intrinsic loss tangent due to two-level fluctuators in the sample. This can be a time-consuming process if one needs to "step" the temperature and wait for it to stabilize before acquiring the scattering parameters (S21 and/or S11) using a VNA. However, by using a Pound-loop one is only limited by the thermal time-constant between the sample and the temperature sensor, allowing for the temperature to be ramped relatively quickly.

\begin{acknowledgments}
The authors would like to thank Phil Meeson, Gregoire Ithier, John Gallop, Ling Hao and Enrico Rubiola for stimulating discussions, advice and help.
This work was supported by EPSRC and the Pathfinder program of the National Measurement Office.
\end{acknowledgments}


\bibliography{pound_res_char}

\end{document}